\documentclass[twocolumn,aps,prl,preprintnumbers]{revtex4}
\usepackage{bbm}
\usepackage{amsmath}
\usepackage{amssymb}
\usepackage{graphicx}
\usepackage{mathrsfs}
\usepackage{amsfonts}
\usepackage{amsthm}
\usepackage{color}
\usepackage{txfonts}
\usepackage[colorlinks=true,citecolor=blue,linkcolor=blue,urlcolor=blue,anchorcolor=blue]{hyperref}%
\usepackage{pifont}
\hypersetup{colorlinks=true,citecolor=blue,linkcolor=blue,urlcolor=blue}
\providecommand{\U}[1]{\protect\rule{.1in}{.1in}}
\makeatletter
\@ifundefined{textcolor}{}
{
\definecolor{BLACK}{gray}{0}
\definecolor{WHITE}{gray}{1}
\definecolor{RED}{rgb}{1,0,0}
\definecolor{GREEN}{rgb}{0,1,0}
\definecolor{BLUE}{rgb}{0,0,1}
\definecolor{CYAN}{cmyk}{1,0,0,0}
\definecolor{MAGENTA}{cmyk}{0,1,0,0}
\definecolor{YELLOW}{cmyk}{0,0,1,0}
}

\makeatother
\begin{document}
\title{Giant mode splitting of azimuthal spin waves in radial vortices}

\author{Zhenyu Wang$^{1}$}
\email[Corresponding author: ]{vcwang@hnu.edu.cn}
\author{Liangrui Li$^{1}$}
\author{Xuejuan Liu$^{2}$}
\author{Xiansi Wang$^{1}$}
\author{Ruifang Wang$^{3}$}
\author{H. Y. Yuan$^{4}$}
\email[Corresponding author: ]{hyyuan@zju.edu.cn}

\affiliation{$^{1}$School of Physics and Electronics, Hunan University, Changsha 410082, China\\
$^{2}$Shenzhen Key Laboratory of Ultraintense Laser and Advanced Material Technology, Center for Intense Laser Application Technology, and College of Engineering Physics, Shenzhen Technology University, Shenzhen 518118, China\\
$^{3}$Department of Physics, Xiamen University, Xiamen 361005, China\\
$^{4}$Institute for Advanced Study in Physics, Zhejiang University, Hangzhou 310027, China}

\begin{abstract}
Radial vortex is a topological spin texture stabilized by the interfacial Dzyaloshinskii-Moriya interaction (DMI) in ferromagnetic disks. Previous investigations have shown that the doublet splitting of azimuthal modes in traditional circular vortices arises from the coupling between azimuthal spin waves and vortex core (VC), an effect occurs only for azimuthal indices $m=\pm1$ and is absent for higher-order modes. Here, we present a giant mode splitting of azimuthal spin waves in radial vortices, even in the absence of the VC. This mode splitting arises from the DMI, which can be an order of magnitude larger than that induced by the VC. Moreover, the DMI-induced frequency splitting increases with both the DMI constant and mode index, reaching tens of GHz for higher-order azimuthal modes. Our results reveal a robust mechanism for mode splitting in chiral magnetic textures and deepen the fundamental understanding of the DMI effect on the spin-wave dynamics in confined magnets.
\end{abstract}

\maketitle
\section{Introduction}\label{sec1}
Magnetic vortices are spontaneously formed spin textures in nanoscale ferromagnetic disks, holding potential applications in magnetic random access memories \cite{Bohlens2008,Nakano2011}, spin-torque nano-oscillators \cite{Pribiag2007,Hrkac2015}, neuromorphic computing \cite{Korber2023,Wu2024}, and biomedicine \cite{Kim2010,Yang2015}. A magnetic vortex consists of a curling in-plane magnetization with closed magnetic flux surrounding a nanoscale region known as vortex core (VC) with out-of-plane magnetization. For clarity, this traditional vortex structure is hereafter referred to as a circular vortex to distinguish it from other vortex types. In the presence of the interfacial Dzyaloshinskii-Moriya interaction (DMI), the in-plane magnetization around the VC changes from azimuthal to radial orientation, giving rise to a new vortex state known as radial vortex \cite{Siracusano2016}.
Furthermore, the DMI can reduce the critical size of vortex-hosting disks \cite{Kwon2013,Hu2024}, which is beneficial for applications in high-density magnetic storage devices.

The vortex spin excitations are of interest from a physical perspective and also hold significant potential for applications.
In circular vortices, three types of spin excitation modes have been identified: The gyrotropic \cite{Guslienko2002,Seeger2025}, azimuthal spin-wave \cite{Park2005,Mruczkiewicz2018}, and radial spin-wave modes \cite{Buess2004,Wang2017}. Interactions between these modes lead to a rich variety of intriguing phenomena, including spin-wave-driven VC switching \cite{Kammerer2011,Wang2012}, doublet splitting of azimuthal spin-wave modes \cite{Park2005,Guslienko2008}, whispering-gallery magnons \cite{Schultheiss2019}, and twisted magnon frequency comb \cite{Wang2022}. Among these, the splitting of azimuthal modes with indices $m=\pm1$ is commonly attributed to the dynamic hybridization between azimuthal spin waves and VC \cite{Park2005,Guslienko2008,Hoffmann2007}. However, recent work has revealed that frequency splitting is not exclusive to $|m|=1$ modes, but can also be observed in higher-order azimuthal modes ($|m|>1$) \cite{Salama2025}. This indicates that the dynamical coupling between azimuthal spin waves and gyrotropic modes is not the sole mechanism for azimuthal mode splitting. For instance, frequency splitting has also been reported in cone vortices under an external magnetic field, even when the VC is removed \cite{Ivanov2002,Ivanov2005,Uzunova2019,Uzunova2023}. These observations motivate us to explore alternative mechanisms underlying the azimuthal mode splitting.

The nonreciprocity of azimuthal spin waves typically originates from the symmetry breaking induced by dipolar interactions and tends to diminish for short-wavelength higher-order modes \cite{Salama2025}. In the exchange-dominated regime, the DMI can induce the spin-wave nonreciprocity, which exhibits two features. One is the frequency difference between counterpropagating spin waves, widely used to quantify the DMI constant \cite{Zakeri2010,Di2015}. The other is the asymmetric wavelength of spin waves at the same frequency \cite{Moon2013}. In confined geometries, this feature manifests as unusual standing waves lacking well-defined nodes \cite{Zingsem2019}. The DMI-induced nonreciprocity depends on the in-plane angle between the spin-wave wavevector and the magnetization \cite{Cortes2013,Zhang2015}, peaking when the wavevector is perpendicular to the magnetization ($\mathbf{k}\bot\mathbf{m}$). In circular vortices, radial spin waves satisfy $\mathbf{k}\perp\mathbf{m}$, and the DMI causes these waves to travel directionally rather than form standing waves \cite{Wang2020,Flores2020}. Similarly, azimuthal spin waves in radial vortices fulfill $\mathbf{k}\bot\mathbf{m}$, providing an ideal platform to investigate the DMI-induced azimuthal mode splitting.

In this work, we study the splitting of azimuthal spin-wave modes in radial vortices. As the DMI increases, the spin configuration of the nanodisk evolves from a circular vortex to a radial vortex.
Meanwhile, the mode splitting of azimuthal spin waves undergoes significant changes. Specifically, the frequency difference in radial vortices is enhanced by approximately one order of magnitude compared to that in circular vortices. By excluding the effects of VC and spin configuration, we attribute this giant mode splitting to the DMI.
Furthermore, this DMI-induced frequency splitting increases with both the DMI constant and mode index, reaching tens of GHz for higher-order azimuthal modes, which can be readily detected in experimental measurements.

\section{Theoretical Model}\label{sec2}
We consider a ferromagnetic nanodisk of radius $R$ and thickness $L$ with a vortex ground state. In polar coordinates, we express the magnetization in terms of spherical polar and azimuthal angles as $\mathbf{m}=\{\sin\theta\cos\varphi,\sin\theta\sin\varphi,\cos\theta\}$. Due to the radial symmetry of the vortex magnetization, the polar and azimuthal angles of the magnetization only depend on $\rho$, i.e., $\theta(\rho)$ and $\varphi(\rho)$. The total magnetic energy of the disk is written as
\begin{equation}\label{eq_Etot}
 E_{\mathrm{tot}}=E_{\mathrm{ex}}+E_{\mathrm{dmi}}+E_{\mathrm{dip}},
\end{equation}
where the contributions from the exchange interaction, interfacial DMI, and dipolar interaction are given respectively by
\begin{subequations}\label{eq_Ei}
\begin{align}
&E_{\mathrm{ex}}=2\pi LA\int\bigg[\bigg(\frac{d\theta}{d\rho}\bigg)^2+\sin^2\theta\bigg(\frac{d\varphi}{d\rho}\bigg)^2+\frac{\sin^2\theta}{\rho^2}\bigg]\rho d\rho,\\
&E_{\mathrm{dmi}}= 2\pi LD\int\biggl[\bigg(\frac{d\theta}{d\rho}+\frac{\sin\theta\cos\theta}{\rho}\bigg)\cos\varphi \\
   &\qquad -\sin\theta\cos\theta\sin\varphi\frac{d\varphi}{d\rho}\biggr]\rho d\rho,\notag \\
&E_{\mathrm{dip}}=\pi L \mu_0 M_{s}^2\int\bigg(\cos^2\theta+\frac{2L}{R}\sin^2\theta\cos^2\varphi\bigg)\rho d\rho.
\end{align}
\end{subequations}
Here, $A$ is the exchange constant, $D$ is the DMI strength, $M_s$ is the saturated magnetization, and $\mu_0$ is the vacuum permeability.
Generally, the dipolar energy is a nonlocal functional of the magnetization, which precludes further quantitative analysis. We therefore retain only the contributions from the surface magnetic charges at the disk's top and bottom surfaces and the local radial demagnetization effect in the radial direction \cite{Verba2020}, resulting in the simplified dipolar energy expressed in Eq. (\ref{eq_Ei}c).

For circular vortices with the DMI, the azimuthal angle ($\varphi$) of the magnetization varies only within the core region ($\rho\ll R$) and remains nearly constant elsewhere \cite{Flores2020}. To simplify calculations, we assume $\partial_{\rho}\varphi=0$. Minimizing the energy function Eq. (\ref{eq_Etot}) with the Euler-Lagrange method, we have
\begin{align}\label{eq_thtfi}
& A\bigg(\frac{d^{2}\theta}{d\rho^{2}}+\frac{1}{\rho}\frac{d\theta}{d\rho}-\frac{\sin\theta\cos\theta}{\rho^2}\bigg)+D\frac{\cos\varphi}{\rho}\sin^2\theta \notag \\
&+\frac{\mu_0M_s^2}{2}(1-\frac{2L}{R}\cos^2\varphi)\sin\theta\cos\theta=0,
\end{align}
with the boundary condition
\begin{equation}\label{eq_BC}
\theta|_{\rho=0}=0, \quad
\frac{d\theta}{d\rho}\Big|_{\rho=R}=\frac{D}{2A}.
\end{equation}

To ascertain the critical DMI for the circular-to-radial vortex phase transition, we calculate the vortex energy in the ($D,\varphi$) parameter space. For each group of $D$ and $\varphi$, the radial profile of the vortex $\theta(\rho)$ can be numerically calculated by using the Runge-Kutta method. Substituting $\theta(\rho)$ into Eqs. (\ref{eq_Etot}) and (\ref{eq_Ei}), the vortex energies are calculated, as plotted in Fig. \ref{fig1}(b). For a given $D$, the minimal energy value corresponds to the ground-state vortex characterized by the associated $\varphi$ [see black dots in Fig. \ref{fig1}(b)]. According to the variation of $\varphi$, we can determine the critical DMI for the vortex phase transition to be $|D_c|=1.35$ $\mathrm{mJ/m^2}$.

In the following, we focus on the radial vortex and investigate the spin-wave excitation on top of this equilibrium magnetization.
Radial vortices exhibit antiradial chirality ($\varphi=\pi$) for $D>0$ and radial chirality ($\varphi=0$) for $D<0$ [see Figs. 1(a) and 1(b)].
The spin-wave dynamics in the vortex state are described by the Landau-Lifshitz-Gilbert (LLG) equation,
\begin{equation}\label{eq_LLG}
  \frac{\partial\mathbf{m}}{\partial t}=-\gamma\mathbf{m}\times\mathbf{H}_{\mathrm{eff}}+\alpha\mathbf{m}\times\frac{\partial\mathbf{m}}{\partial t},
\end{equation}
where $\gamma$ is the gyromagnetic ratio and $\alpha$ is the Gilbert damping constant. The effective field $\mathbf{H}_{\mathrm{eff}}$ includes the exchange, dipolar, and DM fields.
To describe small magnetization oscillations, we introduce a local coordinate system $(\hat{\rho}^{\prime},\hat{\phi}^{\prime},\hat{z}^{\prime})$ such that the equilibrium magnetization $\mathbf{m}_{0}$ always aligns along the $\hat{\rho}^{\prime}$ direction:
\begin{equation}\label{eq_cd}
  \left(
    \begin{array}{c}
      \hat{\rho}^{\prime} \\
      \hat{\phi}^{\prime} \\
      \hat{z}^{\prime} \\
    \end{array}
  \right)=\left(
            \begin{array}{ccc}
              \sin\theta\cos\varphi & 0 & \cos\theta \\
              0 & 1 & 0 \\
              -\cos\theta & 0 & \sin\theta\cos\varphi \\
            \end{array}
          \right)\left(
                   \begin{array}{c}
                     \hat{\rho} \\
                     \hat{\phi} \\
                     \hat{z} \\
                   \end{array}
                 \right).
\end{equation}
The magnetization and effective field are decomposed into static and dynamic parts:
\begin{subequations}\label{eq_mH}
\begin{align}
    \mathbf{m} &= m_{0}\hat{\rho}^{\prime}+m_{\phi^{\prime}}\hat{\phi}^{\prime}+m_{z^{\prime}}\hat{z}^{\prime}, \\
    \mathbf{H}_{\mathrm{eff}} &= H_{\mathrm{eff}}^0\hat{\rho}^{\prime}+h_{\phi^{\prime}}\hat{\phi}^{\prime}+h_{z^{\prime}}\hat{z}^{\prime},
\end{align}
\end{subequations}
with $m_{0}\approx1$ and $m_{\phi^{\prime},z^{\prime}}\ll1$. To obtain the spin-wave dispersion, we define the spin-wave function $\Phi=m_{\phi^{\prime}}-i m_{z^{\prime}}=a_{n}(\rho)\exp[i(m\phi-\omega t)]$, where $a_n(\rho)$ is the radial profile of spin waves, $n$ and $m$ are the radial and azimuthal mode indices, respectively. The dynamic effective field in Eq. (\ref{eq_mH}b) is linearly related to the dynamic magnetization via $h_i=-\Lambda_{ij}m_j$, where $\Lambda_{ij}$ is the magnetic self-interaction tensor.
Neglecting the damping and higher-order terms of the dynamic magnetization ($m_{\phi^{\prime},z^{\prime}}$), the LLG equation Eq. (\ref{eq_LLG}) can be linearized as
\begin{equation}\label{eq_linear_LLG}
  \omega\left(
          \begin{array}{c}
            m_{\phi^{\prime}} \\
            m_{z^{\prime}} \\
          \end{array}
        \right)=i\gamma\left(
                         \begin{array}{cc}
                           -\Lambda_{z^{\prime}\phi^{\prime}} & -\Lambda_{z^{\prime}z^{\prime}}+H_{\mathrm{eff}}^0 \\
                           \Lambda_{\phi^{\prime}\phi^{\prime}}-H_{\mathrm{eff}}^0 & \Lambda_{\phi^{\prime}z^{\prime}} \\
                         \end{array}
                       \right)\left(
          \begin{array}{c}
            m_{\phi^{\prime}} \\
            m_{z^{\prime}} \\
          \end{array}
        \right),
\end{equation}
where
\begin{subequations}\label{eq_component}
\begin{align}
 & H_{\mathrm{eff}}^{0} = -\frac{2A}{M_{s}}\bigg[\bigg(\frac{d\theta}{d\rho}\bigg)^2+\frac{\sin^2\theta}{\rho^2}\bigg]-\frac{2D\cos\varphi}{M_s}\bigg(\frac{d\theta}{d\rho}+
  \frac{\sin\theta\cos\theta}{\rho}\bigg)\notag\\
 &\qquad -\mu_0M_s\bigg(\frac{2L}{R}\sin^2\theta+\cos^2\theta\bigg), \\
 & \Lambda_{\phi^{\prime}\phi^{\prime}} = -\frac{2A}{M_{s}}\bigg(\nabla^2-\frac{1}{\rho^2}\bigg)+\mu_0 M_s\frac{2L}{R},\\
 & \Lambda_{z^{\prime}z^{\prime}} =-\frac{2A}{M_{s}}\bigg[\nabla^2-\bigg(\frac{d\theta}{d\rho}\bigg)^2-\frac{\cos^2\theta}{\rho^2}\bigg]+\frac{2D\cos\varphi}{M_s}\bigg(
 \frac{d\theta}{d\rho}-\frac{\sin\theta\cos\theta}{\rho}\bigg)\notag\\
 &\qquad +\mu_0 M_s\bigg(\frac{2L}{R}\cos^2\theta+\sin^2\theta\bigg),\\
 & \Lambda_{\phi^{\prime}z^{\prime}}=-\Lambda_{z^{\prime}\phi^{\prime}}=i m\bigg(\frac{2A}{M_s}\frac{2\cos\theta}{\rho^2}-\frac{2D\cos\varphi}{M_s}\frac{\sin\theta}{\rho}\bigg).
\end{align}
\end{subequations}
By solving the eigenvalue problem Eq. (\ref{eq_linear_LLG}), the spin-wave dispersion is obtained
\begin{equation}\label{eq_dispersion}
  \omega=\gamma\Bigg[i\tilde{\Lambda}_{\phi^{\prime}z^{\prime}}+\sqrt{(\tilde{\Lambda}_{\phi^{\prime}\phi^{\prime}}+\tilde{H}_{\mathrm{eff}}^{0})
  (\tilde{\Lambda}_{z^{\prime}z^{\prime}}+\tilde{H}_{\mathrm{eff}}^{0})}\Bigg],
\end{equation}
where the tilde denotes that the quantities in Eq. (\ref{eq_component}) are averaged along the disk radius.
The only term in Eq. (\ref{eq_dispersion}) that changes sign under inversion of the azimuthal index is $\tilde{\Lambda}_{\phi^{\prime}z^{\prime}}$. The frequency difference between azimuthal modes with $\pm m$ is then given by
\begin{equation}\label{eq_dw}
  \Delta\omega=\omega_{+m}-\omega_{-m}=-|m|\frac{4\gamma}{M_s}\bigg[2A\Big\langle\frac{\cos\theta}{\rho^2}\Big\rangle-D\cos\varphi\Big\langle\frac{\sin\theta}{\rho}\Big\rangle\bigg].
\end{equation}
An exact analytical solution for $\theta(\rho)$ in Eq. (\ref{eq_thtfi}) is unavailable, making precise calculation of $\Delta\omega$ difficult. Nevertheless, from Eq. (\ref{eq_dw}) we can qualitatively conclude  that the frequency difference $\Delta\omega$ increases with the DMI constant $D$ and azimuthal index $|m|$.

\section{Micromagnetic simulations}

To verify our theoretical analysis, micromagnetic simulations are performed using the MuMax3 code \cite{Vansteenkiste2014}. In simulations, a ferromagnetic disk with a 100 nm diameter and 1 nm thickness is discretized by the cell size of $1\times1\times1$ $\mathrm{nm}^3$. Magnetic parameters of permalloy (Py) are adopted: the exchange constant $A_{\mathrm{ex}}=13$ $\mathrm{pJ/m}$, saturation magnetization $M_s=8.6\times10^5$ $\mathrm{A/m}$, and the damping constant $\alpha=0.01$.
A significant DMI can be induced at the interface between a Py layer and heavy metals \cite{Nembach2015,Kuepferling2023} or two-dimensional materials \cite{Chaurasiya2019,Vas2024} via symmetry breaking with strong spin-orbit coupling. By modifying the stacking order of the Py and heavy metal layers, the sign of DMI can be reversed \cite{Cho2017}.

We first simulate the static magnetization configuration of the nanodisk for DMI constants ranging from -3 to 3 $\mathrm{mJ/m^2}$. A circular vortex with polarity $p=1$ and chirality $c=1$ is set as the initial state and relaxed to the ground state. The stable magnetization states are then excited by a sinc-function field with spatial modulation, which has the following form \cite{Korber2020}:
\begin{equation}\label{eq_sinc}
  \mathbf{h}(\phi,t)=h_0 \cos(m\phi) \frac{\sin[\omega_c(t-t_0)]}{\omega_c(t-t_0)} \hat{z},
\end{equation}
with $h_0=5$ mT, $\omega_c/2\pi=100$ GHz, $t_0=10$ ns, and the total simulation time of 100 ns. The spatially modulated function $\cos(m\phi)$ is chosen for the efficient excitation of azimuthal spin waves with the mode index $\pm m$. By performing fast Fourier transformation (FFT) of the dynamical magnetization $\delta m_z$, we can obtain the spin excitation spectrum for extracting eigenfrequencies and their corresponding mode profiles.

\section{Results}
\subsection{Circular-to-radial vortex phase transition}

\begin{figure}
  \centering
  \includegraphics[width=1\linewidth]{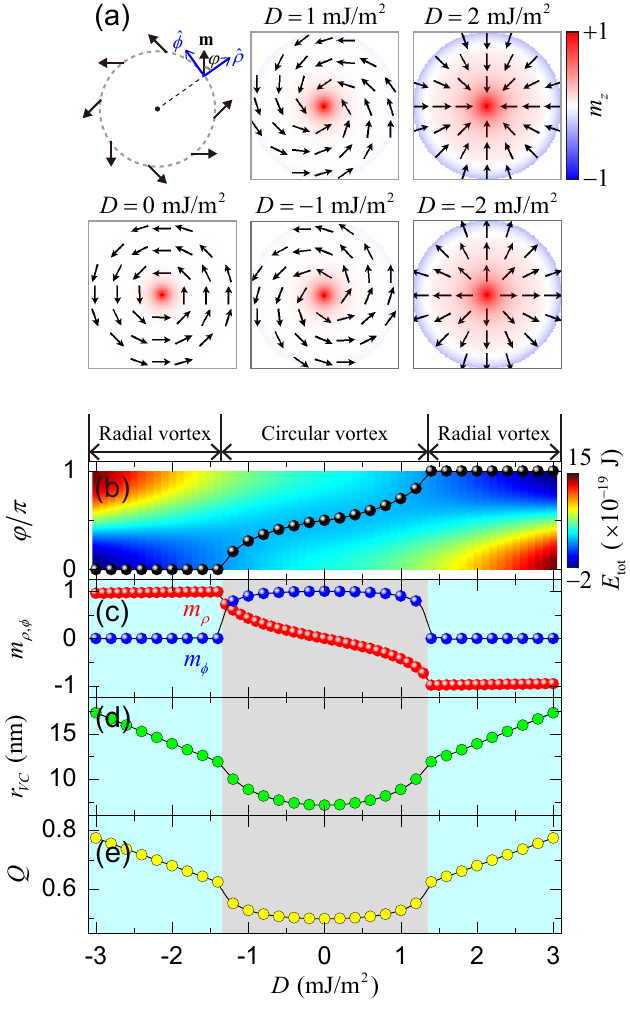}\\
  \caption{(a) Static spin configurations of the vortices for $D=0$, $\pm1$, and $\pm2$ $\mathrm{mJ/m^2}$. Blue arrows depict the two dimensional cylindrical coordinate system ($\hat{\rho},\hat{\phi}$). Black arrows show the in-plane magnetization and the color bar represents the out-of-plane magnetization. (b) The vortex energy in $(D,\varphi)$ space, black symbols represent the associated $\varphi$ for minimum energy at each $D$. (c) The magnetization components ($m_{\rho,\phi}$) at $\rho=25$ nm, (d) the radius of the vortex core ($r_{VC}$), and (e) topological charge ($Q$) versus the DMI constant ($D$).}\label{fig1}
\end{figure}

Figure \ref{fig1}(a) shows five representative spin configurations of the vortices under different DMI values. At $D=0$ $\mathrm{mJ/m^2}$, a circular vortex is stabilized by the balance between the exchange and dipolar interactions. When the DMI is introduced ($|D|=1$ $\mathrm{mJ/m^2}$), the in-plane magnetization tilts from the azimuthal to radial directions. A radial vortex is formed when the DMI is sufficiently strong to overcome the dipolar interaction ($|D|=2$ $\mathrm{mJ/m^2}$).
By increasing the DMI strength ($|D|$) from 0 to 3 $\mathrm{mJ/m^2}$ with a step of 0.1 $\mathrm{mJ/m^2}$, we extract the magnetization components at $\rho=25$ nm.
The azimuthal component ($m_\phi$) of the magnetization decreases from 1 to 0, while the radial component ($m_\rho$) varies from 0 to $\mp1$ [see the blue and red dots in Fig. \ref{fig1}(c)]. At $|D|=1.4$ $\mathrm{mJ/m^2}$, the azimuthal component vanishes and radial component changes to $\mp1$, corresponding to the circular-to-radial vortex phase transition. The critical DMI for the vortex phase transition obtained from simulations agrees well with the analytical result shown in Fig. \ref{fig1}(b).
It is noted that the critical DMI increases with the disk thickness due to the enhanced dipolar interaction (see Appendix \ref{Appendix_A}).

Besides the magnetization configuration, we also study the DMI effect on the VC size. To extract the VC size, the magnetization along the disk diameter is fitted by the Walker profile, which is commonly utilized to describe the spin configuration of domain wall \cite{Schryer1974}. It is found that the VC radius increases with the DMI strength ($|D|$), reaching 17.5 nm (larger than one-third of the disk radius) at $|D|=3$ $\mathrm{mJ/m^2}$ [see the green dots in Fig. \ref{fig1}(d)].
This trend is confirmed by the out-of-plane component ($m_z$) of the vortex spin configuration for different DMIs (see Appendix \ref{Appendix_B}). Additionally, it is observed that vortices with opposite DMI signs exhibit identical $m_z$ (i.e., $\theta$) profiles.
The topological charge of the system is calculated by using the formula $Q=(1/4\pi)\iint\mathbf{m}\cdot(\partial_x\mathbf{m}\times\partial_y\mathbf{m})dxdy$. The topological charge increases from 0.5 at $D=0$ $\mathrm{mJ/m^2}$ to 0.78 at $|D|=3$ $\mathrm{mJ/m^2}$ [see the yellow dots in Fig. \ref{fig1}(e)], which is due to the DMI-induced spin canting at the disk edge \cite{Mulkers2017}. An abrupt change of $Q$ is observed at the phase transition point $|D|=1.35$ $\mathrm{mJ/m^2}$.

\subsection{Azimuthal mode splitting}
Next, we focus on the azimuthal mode splitting of the vortex state for $D>0$. Complementary results for the case of $D<0$ are provided in Appendix \ref{Appendix_C}.
Figure \ref{fig2}(a) shows a continuous frequency evolution of azimuthal spin waves with $m=\pm 1$ across the circular-to-radial vortex phase transition. In circular vortices, the eigenfrequencies of both modes decrease with $D$, with the frequency of the $m=+1$ mode decreasing faster than that of the $m=-1$ mode. The frequency branches of two modes intersect at $D=1.1$ $\mathrm{mJ/m^2}$. In radial vortex states, the eigenfrequencies of both azimuthal modes increase with $D$, with the $m=+1$ mode increasing more slowly.

The mode profiles of azimuthal spin waves at three different DMI constants are plotted in Fig. \ref{fig2}(b). At $D=0$ $\mathrm{mJ/m^2}$, it corresponds to the traditional circular vortex, a topic extensively explored in prior researches \cite{Park2005,Mruczkiewicz2018}. A mode node appears near the disk center, which arises from the coupling between azimuthal modes and VC gyration. The spatial extent and amplitude inside the node reflect the coupling strength. The area within the node of the $m=+1$ mode is larger than that of the $m=-1$ mode, indicating a stronger coupling between the VC gyrotropic mode and $m=+1$ mode compared to the $m=-1$ mode.
This has been confirmed by the fact that the VC gyration is more easily driven by the $m=+1$ mode \cite{Kammerer2011,Wang2022}.
Moreover, the wavefront of azimuthal spin waves is along the radial direction, corresponding to the wavevector parallel with the azimuthal direction ($\mathbf{k}\parallel\hat{\phi}$).

At $D=0.5$ $\mathrm{mJ/m^2}$, the vortex state retains its circular form. Although such a weak DMI is insufficient to alter the static configuration, it does affect the vortex spin excitation.
In ferromagnetic films with uniform in-plane magnetization, the DMI shifts the isofrequency curve of spin waves, resulting in noncollinearity between the wavevector and group velocity ($\mathbf{v}_g$) when the wavevector is not perpendicular to the magnetization \cite{Wang2018}.
For $\mathbf{m}\parallel\mathbf{v}_g$, it manifests as spin-wave canting \cite{Guo2017}. A similar phenomenon occurs in circular vortices with the DMI where $\mathbf{v}_g\parallel\mathbf{m}$. The DMI tilts the wavevector from the azimuthal direction, twisting the wavefront of azimuthal spin waves into a spiral pattern [see Fig. \ref{fig2}(b)].

A radial vortex stabilizes at $D=2$ $\mathrm{mJ/m^2}$. In this configuration, azimuthal spin waves meet the condition of $\mathbf{k}\perp\mathbf{m}$, resulting in a significant frequency splitting of the $m=\pm 1$ modes. According to the mode amplitudes within the node, the $m=+1$ mode couples much more strongly to the gyrating VC than the $m=-1$ mode, consistent with the observation that the VC gyration strongly shifts the $m=+1$ mode frequency but barely affects the $m=-1$ mode (not shown here).

\begin{figure}
  \centering
  \includegraphics[width=1\linewidth]{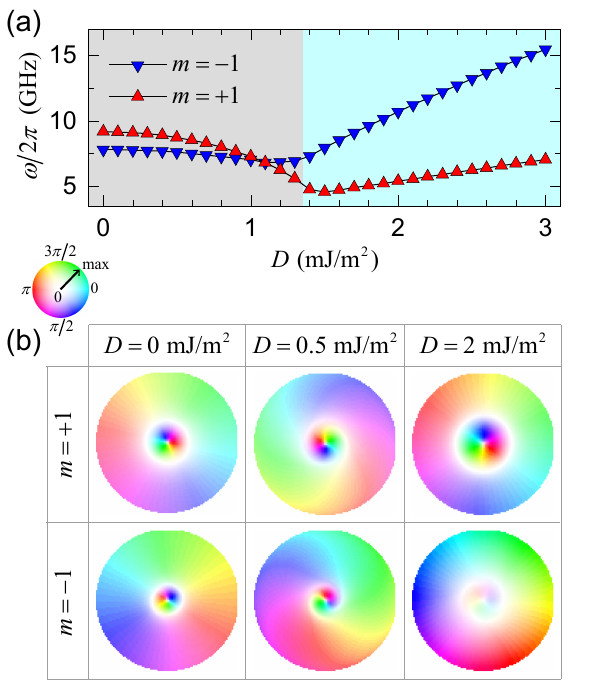}\\
  \caption{(a) The frequencies of azimuthal spin-wave modes with $m=\pm1$ versus the DMI constant. (b) The mode profiles of azimuthal spin waves with $m=\pm1$ for $D=0$, 0.5, and 2 $\mathrm{mJ/m^2}$. The inset is the color palette of azimuthal spin-wave modes with color indicating the phase and intensity representing the amplitude.}\label{fig2}
\end{figure}

Figure \ref{fig3}(a) shows the frequency splitting ($\Delta\omega$) of azimuthal modes as a function of $D$. One can see that $\Delta\omega$ is changed from 1.36 GHz at $D=0$ $\mathrm{mJ/m^2}$ to $-8.46$ GHz at $D=3$ $\mathrm{mJ/m^2}$. In radial vortices, the magnitude of the frequency splitting ($|\Delta\omega|$) increases linearly with $D$, consistent with Eq. (\ref{eq_dw}), and is much larger than that in circular vortices.
In extended ferromagnetic films, the DMI-induced frequency shift scales as $2Dk\sin\psi$ \cite{Di2015}, where $k$ is the wave number of spin waves and $\psi$ is the angle between $\mathbf{k}$ and $\mathbf{m}$. This relation holds true in confined disks as well.
For circular vortices ($\psi\approx0$), the DMI effect can be disregarded, and the splitting is dominated by the VC gyration. For radial vortices ($\psi=\pi/2$), both the VC and DMI contribute to the mode splitting. Therefore, it is essential to ascertain their relative contributions.

\subsection{The DMI effect on mode splitting}

\begin{figure}
  \centering
  \includegraphics[width=1\linewidth]{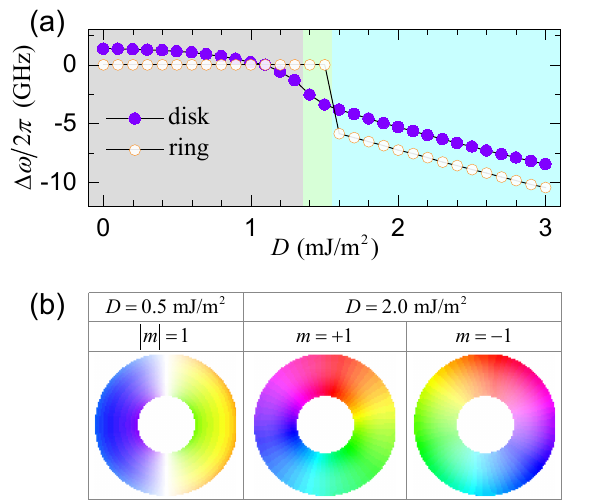}\\
  \caption{(a) The frequency splitting as a function of the DMI constant for the ferromagnetic disk and ring. The inner radius of the ring is 20 nm. (b) The mode profiles of azimuthal spin waves in the ring for $D=0.5$ and 2.0 $\mathrm{mJ/m^2}$.}\label{fig3}
\end{figure}

To assess the DMI contribution, we examine the spin excitations of azimuthal modes in a ferromagnetic ring featuring a central circular hole of 40 nm diameter, which eliminates the VC effect. In this ferromagnetic ring, the circular vortex remains stable up to $D=1.5$ $\mathrm{mJ/m^2}$, exceeding the stabilization range in full disks [see the light green area in Fig. \ref{fig3}(a)]. This enhanced stability is attributed to the dipolar interactions in the ring geometry. In circular vortex rings ($0\leq D\leq1.5$ $\mathrm{mJ/m^2}$), the mode splitting vanishes ($\Delta\omega=0$) [see orange open circles in Fig. \ref{fig3}(a)], as the VC is removed and the DMI does not induce a frequency splitting when $\mathbf{k}\parallel\mathbf{m}$. This would lead to a mode degeneracy, characterized by an azimuthal nodal line [see the left panel ($D=0.5$ $\mathrm{mJ/m^2}$) of Fig. \ref{fig3}(b)]. Moreover, it is noted that the nonuniform phase along the radial direction arises from the superposition of azimuthal spin waves with spiral wavefronts induced by the DMI.

In radial vortex rings, the mode splitting of azimuthal spin waves with $m=\pm1$ persists even without the VC [see Figs. \ref{fig3}(a) and \ref{fig3}(b)], suggesting that the giant mode splitting in radial vortices originates mainly from the DMI effect.
Sweeping the central hole diameter across a broad range verifies that annular boundary effects are negligible, and DMI-mediated splitting remains robust against geometric modification (see Appendix \ref{Appendix_D}).
Furthermore, the frequency asymmetry is more pronounced compared to the disks containing the VC, indicating the VC and DMI contribute oppositely to the mode splitting. The VC gyration results in a positive frequency difference (i.e. $\Delta\omega>0$), whereas the DMI yields $\Delta\omega<0$. In radial vortices, the polarity and chirality are locked due to the DMI-induced energy nondegeneracy \cite{Siracusano2016}. Reversing the VC polarity to $p=-1$ switches both contributions from both the VC and DMI, making the DMI-induced splitting positive and VC-induced splitting negative.
This phenomenon is fully explained by the generalized mode splitting formula [Eq. (\ref{eq_dw})]. It also allows us to infer that for fixed vortex polarity, radial vortices with opposite DMI signs exhibit the identical frequency splitting, as confirmed by the simulation results in Appendix \ref{Appendix_C}.

In circular vortices, the DMI tilts in-plane magnetization from the azimuthal to radial direction [see Fig. \ref{fig1}(b)], enhancing the DMI contribution to the mode splitting. As $D$ increases, it neutralizes the VC contribution, resulting in $\Delta\omega=0$ at $D=1.1$ $\mathrm{mJ/m^2}$, as shown in Fig. \ref{fig3}(a). This also corresponds to the crossing of two frequency branches at $D=1.1$ $\mathrm{mJ/m^2}$ in Fig. \ref{fig2}(a).

\subsection{Effect of the spin configuration}

\begin{figure}
  \centering
  \includegraphics[width=1\linewidth]{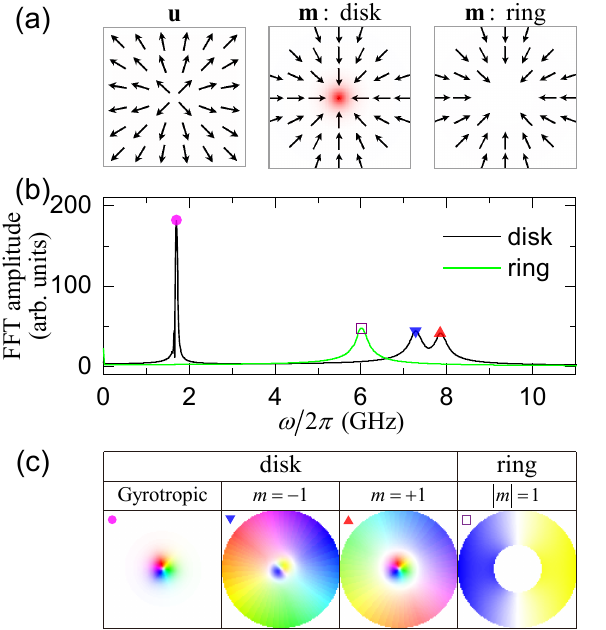}\\
  \caption{(a) The easy-axis direction of the radial magnetic anisotropy and the magnetization configuration in the disk and ring, which are of the same size as those in Fig. \ref{fig3}. (b) The spin excitation spectra of the disk and ring. The magenta dot and blue and red triangles represent the gyrotropic, and azimuthal spin-wave modes with $m=\pm1$, respectively. The purple square denotes the degenerate mode of azimuthal spin waves with $|m|=1$ in the ring. (c) The corresponding mode profiles in (b).}\label{fig4}
\end{figure}

In the preceding subsection, we attribute the mode splitting in radial vortices to the DMI effect after excluding the VC contribution. However, a potential loophole remains in this argument: The spin configuration itself changes from a circular vortex ($D=0$ $\mathrm{mJ/m^2}$) to a radial vortex ($D>1.35$ $\mathrm{mJ/m^2}$), which might account for the observed mode splitting rather than the DMI effect.

To rule out this possibility, we stabilize a radial vortex by radial magnetic anisotropy \cite{Mimica2025} instead of the DMI. The anisotropy field is defined as $\mathbf{H}_{\mathrm{an}}=(2K_{\rho}/M_s)m_{\rho}\hat{\rho}$ with the radial anisotropy constant $K_{\rho}=3\times10^4$ $\mathrm{J/m^3}$ and easy-axis direction $\mathbf{u}=\hat{\rho}$. As shown in Fig. \ref{fig4}(a), both the full disk and ring can form a radial vortex. In the full disk, the spin excitation spectrum shows three peaks corresponding to the gyrotropic, and azimuthal spin-wave modes with $m=\pm1$ [see Fig. \ref{fig4}(b)]. By removing the VC, the gyrotropic mode disappears and azimuthal spin-wave modes with $m=\pm1$ become degenerate. An azimuthal nodal line is observed in the mode profile of azimuthal modes, with a uniform phase along the radial direction [see Fig. \ref{fig4}(c)]. This indicates that the mode splitting in the disk is caused by the VC, not by the vortex spin configuration. We thus conclude that the giant mode splitting in radial vortices is dominated by the DMI effect.

\subsection{Dependence of the frequency splitting on the mode index}
Having clarified the mechanism for the mode splitting with $m=\pm1$, we extend our investigation to higher-order azimuthal modes. To achieve higher-order azimuthal modes, the disk size is increased to a 200 nm diameter and 2 nm thickness, maintaining the same aspect ratio as the above studied disk. We consider a DMI constant $D=3.0$ $\mathrm{mJ/m^2}$, which can stabilize a radial vortex in the disk.

\begin{figure}
  \centering
  \includegraphics[width=1\linewidth]{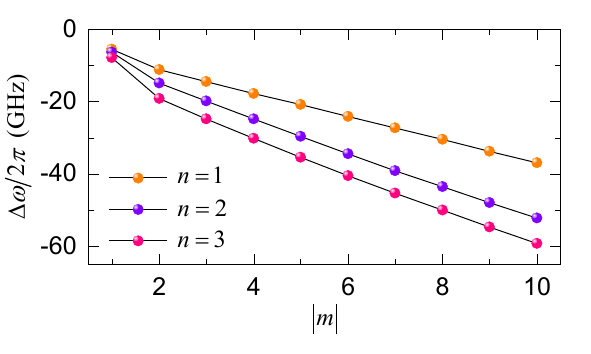}\\
  \caption{The frequency splitting as a function of the azimuthal index ($|m|$) for different radial indices $n$, for a disk with a diameter of 200 nm, thickness of 2 nm, and DMI constant $D=3.0$ $\mathrm{mJ/m^2}$.}\label{fig5}
\end{figure}

Figure \ref{fig5} shows that $|\Delta\omega|$ increases with the azimuthal mode index $|m|$, consistent with Eq. (\ref{eq_dw}).
This finding differs from the VC-induced mode splitting, which vanishes for higher-order modes \cite{Salama2025,Ivanov2002,Ivanov2005}.
Additionally, the frequency splitting also increases with the radial index $n$. For $n=3$ and $|m|=10$, the frequency asymmetry reaches $|\Delta\omega|/2\pi\approx60$ GHz, which can be further increased by enhancing the DMI strength or mode index.

\section{Discussion and Conclusion}

The interfacial DMI is considered here and its induced frequency asymmetry of spin waves reaches the maximum when $\mathbf{k}\perp\mathbf{m}$. This condition is satisfied by azimuthal spin waves in radial vortices, thereby leading to a giant splitting of azimuthal modes. However, for the bulk DMI, the frequency difference peaks at $\mathbf{k}\parallel\mathbf{m}$ \cite{Cortes2013,Yu2016}, which is met by azimuthal spin waves in circular vortices. Therefore, our results are expected to extend to ferromagnetic disks with the bulk DMI.

In summary, we present a giant mode splitting of azimuthal spin waves in radial vortices. By excluding the effect of the VC and spin configuration, we demonstrate that this large mode splitting originates from the DMI effect. Unlike the VC-induced splitting, which is restricted to the $m=\pm1$ modes, the DMI-induced splitting occurs for all azimuthal modes with $|m|\neq0$, and becomes more pronounced. The frequency difference increases with the DMI constant and mode index, which can reach up to tens of GHz for higher-order azimuthal modes. Our findings deepen the fundamental understanding of the DMI effect on spin excitations in confined magnets.

\section*{ACKNOWLEDGMENT}
We thank X. Liang for helpful discussions.
This work is supported by the Fundamental Research Funds for the Central Universities. Z.W. acknowledges the support of the Natural Science Foundation of Hunan Province of China (Grant No. 2024JJ6113) and the Natural Science Foundation of China (NSFC) (Grant No. 12204089).
H.Y. Yuan is supported by the National Key R$\&$D Program of China (2022YFA1402700) and the NSFC (Grant No. 12574132).
X.S.W. acknowledges the support from the NSFC (Grants No. 12174093) and the Natural Science Foundation of Hunan Province of China (Grant No. 2025JJ60001).

\appendix

\section{Effect of the disk thickness}\label{Appendix_A}
\begin{figure}
  \centering
  \includegraphics[width=1\linewidth]{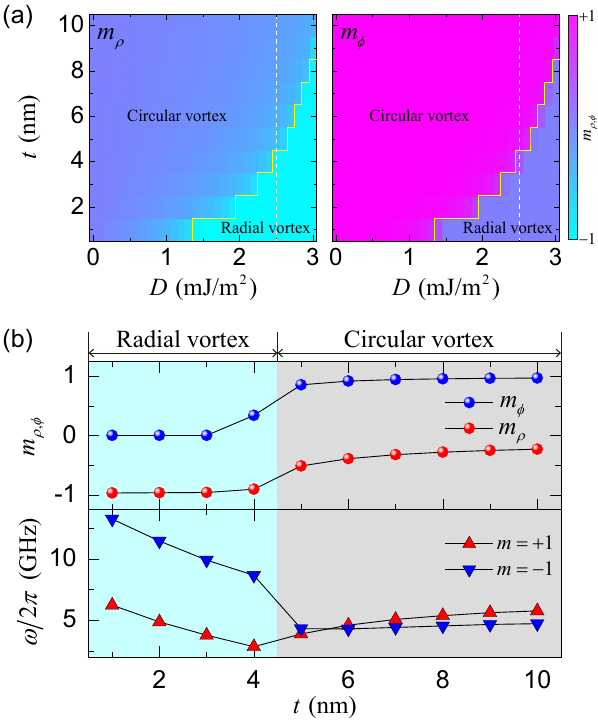}\\
  \caption{(a) The radial and azimuthal components ($m_{\rho,\phi}$) of the vortex magnetization at $\rho=25$ nm in ($D,t$) space. The yellow lines represent the critical DMI constants for the circular-to-radial vortex phase transition. (b) The in-plane components of the vortex magnetization (upper panel) and azimuthal spin-wave frequencies (lower panel) as a function of the disk thickness under a fixed DMI ($D=2.5$ $\mathrm{mJ/m^2}$), labeled by the white dashed lines in (a).}\label{figA1}
\end{figure}

The critical DMI constant for the circular-to-radial vortex phase transition increases with the disk thickness [see Fig. \ref{figA1}(a)]. For $D=2.5$ $\mathrm{mJ/m^2}$, the ground state of the disk is a radial vortex for $t\leq4$ nm, where the DMI dominates the dipolar interaction [see the upper panel of Fig. \ref{figA1}(b)].  As the disk thickness increases ($t>4$ nm), the dipolar interaction becomes stronger and dominates the DMI, resulting in a circular vortex as the ground state. Accordingly, the frequency splitting of the $m=\pm1$ modes is about 6 GHz for radial vortices, much larger than that for circular vortices, as shown in the lower panel of Fig. \ref{figA1}(b). This indicates that the giant splitting of azimuthal modes in radial vortices observed in our work remains robust even with varying the disk thickness.

\section{The out-of-plane magnetization profile of the vortex for different DMIs}\label{Appendix_B}
\begin{figure}
  \centering
  \includegraphics[width=1\linewidth]{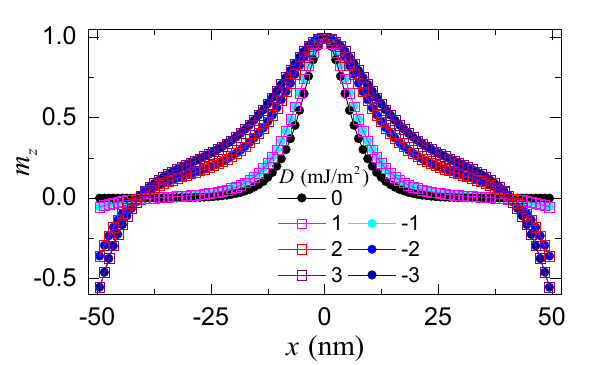}\\
  \caption{The $m_z$ profiles of the vortex along the diameter for $D=0$, $\pm1$, $\pm2$ $\mathrm{mJ/m^2}$.}\label{figB1}
\end{figure}

The out-of-plane magnetization components of the vortex along the disk diameter for seven DMI constants are plotted in Fig. \ref{figB1}. It is found that the $m_z$ profile broadens with increasing the DMI strength ($|D|$), which corresponds to an increase in the vortex core size. Additionally, the DMI-induced tilting angle at the disk edge increases. Notably, vortices with opposite DMI signs exhibit identical $m_z$ (i.e., the polar angle $\theta$) profiles.

\section{Effects of the vortex polarity and DMI sign}\label{Appendix_C}
\begin{figure}
  \centering
  \includegraphics[width=1\linewidth]{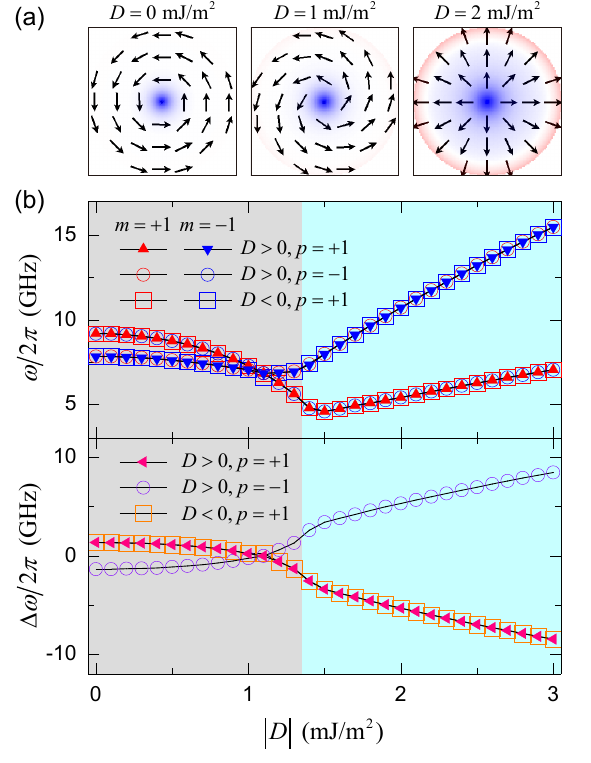}\\
  \caption{(a) Static spin configurations of the vortices with $p=-1$ for $D=0$, 1, 2 $\mathrm{mJ/m^2}$. (b) The azimuthal spin-wave frequencies (upper panel) and their difference (lower panel) as a function of the DMI strength ($|D|$) for three different combinations of ($D,p$).}\label{figC1}
\end{figure}

The DMI lifts the energy degeneracy of the vortex state, resulting in the binding of its polarity and chirality \cite{Siracusano2016}. When the vortex polarity switches from $p=+1$ to $p=-1$, its chirality is also reversed from the antiradial to radial direction [see Figs. \ref{fig1}(a) and \ref{figC1}(a)]. This would lead to the frequency interchange of azimuthal modes (i.e., $\omega_{+m}^{p=-1}=\omega_{-m}^{p=+1}$) \cite{Kammerer2011}, which is confirmed by simulation results [see the upper panel of Fig. \ref{figC1}(b)]. Then, the mode splitting can be written as
\begin{equation}\label{eq_dw2}
  \Delta\omega_{D>0}^{p=-1}=\omega_{+m}^{p=-1}-\omega_{-m}^{p=-1}=\omega_{-m}^{p=+1}-\omega_{+m}^{p=+1}=-\Delta\omega_{D>0}^{p=+1},
\end{equation}
which agrees well with simulation results in the lower panel of Fig. \ref{figC1}(b).
Based on the mode splitting formula Eq. (\ref{eq_dw}), we can draw the same conclusion.
For $D>0$ and $p=+1$, the azimuthal angle of a radial vortex is $\varphi=\pi$, and the mode splitting is given as
\begin{equation}\label{eq_dw3}
  \Delta\omega_{D>0}^{p=+1}=-|m|\frac{4\gamma}{M_s}\bigg[2A\Big\langle\frac{\cos\theta}{\rho^2}\Big\rangle+D\Big\langle\frac{\sin\theta}{\rho}\Big\rangle\bigg].
\end{equation}
When the vortex polarity switches from $p=+1$ to $p=-1$, the polar and azimuthal angles become $\pi-\theta$ and $0$, respectively. Then, we have
\begin{align}\label{eq_dw4}
  \Delta\omega_{D>0}^{p=-1} &= -|m|\frac{4\gamma}{M_s}\bigg[2A\Big\langle\frac{\cos(\pi-\theta)}{\rho^2}\Big\rangle-D\Big\langle\frac{\sin(\pi-\theta)}{\rho}\Big\rangle\bigg] \notag \\
   &= -|m|\frac{4\gamma}{M_s}\bigg[-2A\Big\langle\frac{\cos\theta}{\rho^2}\Big\rangle-D\Big\langle\frac{\sin\theta}{\rho}\Big\rangle\bigg]
   \notag \\
   &=-\Delta\omega_{D>0}^{p=+1},
\end{align}
which is consistent with Eq. (\ref{eq_dw2}).
For $D<0$ and $p=+1$, the reversal of the DMI sign leads to the azimuthal angle changing from $\varphi=\pi$ to $\varphi=0$, with the polar angle $\theta$ remaining unchanged [see Figs. \ref{fig1}(a) and \ref{figB1}]. Then, the mode splitting is calculated as
\begin{align}\label{eq_dw5}
  \Delta\omega_{D<0}^{p=+1}&=-|m|\frac{4\gamma}{M_s}\bigg[2A\Big\langle\frac{\cos\theta}{\rho^2}\Big\rangle-D\Big\langle\frac{\sin\theta}{\rho}\Big\rangle\bigg]
  \notag \\
   &=-|m|\frac{4\gamma}{M_s}\bigg[2A\Big\langle\frac{\cos\theta}{\rho^2}\Big\rangle+|D|\Big\langle\frac{\sin\theta}{\rho}\Big\rangle\bigg]
  \notag \\
   &=\Delta\omega_{D>0}^{p=+1},
\end{align}
which indicates that the vortices with a fixed polarity and opposite DMI signs have the same mode splitting, as confirmed by simulation results in Fig. {\ref{figC1}(b)}.

\section{Effect of the hole diameter}\label{Appendix_D}
\begin{figure}
  \centering
  \includegraphics[width=1\linewidth]{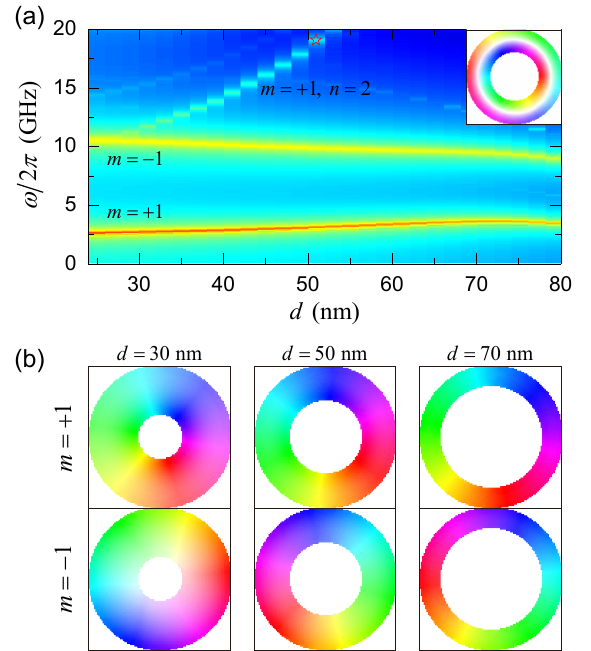}\\
  \caption{(a) The spin-wave spectrum of the ferromagnetic ring with a radial vortex as a function of the hole diameter. The outer diameter is fixed at 100 nm, and the DMI constant is set to $D=2$ $\mathrm{mJ/m^2}$. The inset shows the profile of the $m=+1,n=2$ mode for the hole diameter $d=50$ nm, labeled by the red star in (a). (b) Mode profiles of the azimuthal spin-wave modes ($m=\pm1$) for $d=30,50$ and 70 nm.}\label{figD1}
\end{figure}

To remove the influence of vortex core on the spin-wave spectrum, a ferromagnetic ring with a central hole is employed. The central hole introduces an additional geometry boundary, which may modify the mode frequency and profile. To evaluate the influence of the central hole on the DMI-induced splitting, micromagnetic simulations are performed  with the hole diameter ranging from $d=24$ nm to $d=80$ nm. The results are plotted in Fig. \ref{figD1}, showing that the eigenfrequencies of the $m=\pm1$ modes vary slightly with the hole diameter, and their mode profiles remain nearly unchanged. These results suggest that the DMI-induced splitting is robust against variations in the hole diameter. For high-order modes ($m=+1,n=2$), the hole diameter exerts a significant influence on the mode frequencies, an effect that might be attributed to the occurrence of radial nodes [see the inset of Fig. \ref{figD1}(a)] and warrants further investigation.



\begin{thebibliography}{99}
\bibitem{Bohlens2008} S. Bohlens, B. Kr\"{u}ger, A. Drews, M. Bolte, G. Meier, and D. Pfannkuche, Current controlled random-access memory based on magnetic vortex handedness, \href{https://doi.org/10.1063/1.2998584}{Appl. Phys. Lett. \textbf{93}, 142508 (2008)}.
\bibitem{Nakano2011} K. Nakano, D. Chiba, N. Ohshima, S. Kasai, T. Sato, Y. Nakatani, K. Sekiguchi, K. Kobayashi, and T. Ono, All-electrical operation of magnetic vortex core memory cell, \href{https://doi.org/10.1063/1.3673303}{Appl. Phys. Lett. \textbf{99}, 262505 (2011)}.
\bibitem{Pribiag2007} V. S. Pribiag, I. N. Krivorotov, G. D. Fuchs, P. M. Braganca, O. Ozatay, J. C. Sankey, D. C. Ralph, and R. A. Buhrman, Magnetic vortex oscillator driven by d.c. spin-polarized current, \href{https://doi.org/10.1038/nphys619}{Nat. Phys. \textbf{3}, 498 (2007)}.
\bibitem{Hrkac2015} G. Hrkac, P. S. Keatley, M. T. Bryan, and K. Butler, Magnetic vortex oscillators, \href{https://doi.org/10.1088/0022-3727/48/45/453001}{J. Phys. D: Appl. Phys. \textbf{48}, 453001 (2015)}.
\bibitem{Korber2023} L. K\"{o}rber, C. Heins, T. Hula, J.-V. Kim, S. Thlang, H. Schultheiss, J. Fassbender, and K. Schultheiss, Pattern recognition in reciprocal space with a magnon-scattering reservoir, \href{https://doi.org/10.1038/s41467-023-39452-y}{Nat. Commun. \textbf{14}, 3954 (2023)}.
\bibitem{Wu2024} Y. Wu, Y. Luo, L. Zhang, S. Dai, B. Zhang, Y. Zhou, B. Fang, and Z. Zeng, Adjustable artificial neuron based on vortex magnetic tunnel junction, \href{https://doi.org/10.1063/5.0195602}{Appl. Phys. Lett. \textbf{124}, 122408 (2024)}.
\bibitem{Kim2010} D.-H. Kim, E. A. Rozhkova, I. V. Ulasov, S. D. Bader, T. Rajh, M. S. Lesniak, and V. Novosad, Biofunctionalized magnetic-vortex microdiscs for targeted cancer-cell destruction, \href{https://doi.org/10.1038/nmat2591}{Nat. Mater. \textbf{9}, 165 (2010)}.
\bibitem{Yang2015} Y. Yang, X. Liu, Y. Lv, T. S. Herng, X. Xu, W. Xia, T. Zhang, J. Fang, W. Xiao, and J. Ding, Orientation mediated enhancement on magnetic hyperthermia of $\mathrm{Fe_{3}O_{4}}$ nanodisc, \href{https://doi.org/10.1002/adfm.201402764}{Adv. Funct. Mater. \textbf{25}, 812 (2015)}.
\bibitem{Siracusano2016} G. Siracusano, R. Tomasello, A. Giordano, V. Puliafito, B. Azzerboni, O. Ozatay, M. Carpentieri, and G. Finocchio, Magnetic Radial Vortex Stabilization and Efficient Manipulation Driven by the Dzyaloshinskii-Moriya Interaction and Spin-Transfer Torque, \href{https://doi.org/10.1103/PhysRevLett.117.087204}{Phys. Rev. Lett. \textbf{117}, 087204 (2016)}.
\bibitem{Kwon2013} H. Y. Kwon, S. P. Kang, Y. Z. Wu, and C. Won, Magnetic vortex generated by Dzyaloshinskii-Moriya interaction, \href{https://doi.org/10.1063/1.4799401}{J. Appl. Phys. \textbf{113}, 133911 (2013)}.
\bibitem{Hu2024} X. Hu, X. S. Wang, and Z. Wang, Stabilization and dynamics of magnetic antivortices in a nanodisk with anisotropic Dzyaloshinskii-Moriya interaction, \href{https://doi.org/10.1103/PhysRevB.110.014437}{Phys. Rev. B \textbf{110}, 014437 (2024)}.
\bibitem{Guslienko2002} K. Yu. Guslienko, B. A. Ivanov, V. Novosad, Y. Otani, H. Shima, and K. Fukamichi, Eigenfrequencies of vortex state excitations in magnetic submicron-size disks, \href{https://doi.org/10.1063/1.1450816}{J. Appl. Phys. \textbf{91}, 8037 (2002)}.
\bibitem{Seeger2025} R. Lopes Seeger, F. Millo, G. Soares, J.-V. Kim, A. Solignac, G. de Loubens, and T. Devolder, Experimental Observation of Vortex Gyration Excited by Surface Acoustic Waves, \href{https://doi.org/10.1103/PhysRevLett.134.176704}{Phys. Rev. Lett. \textbf{134}, 176704 (2025)}.
\bibitem{Park2005} J. P. Park and P. A. Crowell, Interactions of Spin Waves with a Magnetic Vortex, \href{https://doi.org/10.1103/PhysRevLett.95.167201}{Phys. Rev. Lett. \textbf{95}, 167201 (2005)}.
\bibitem{Mruczkiewicz2018} M. Mruczkiewicz, P. Gruszecki, M. Krawczyk, and K. Y. Guslienko, Azimuthal spin-wave excitations in magnetic nanodots over the soliton background: Vortex, Bloch, and N\'{e}el-like skyrmions, \href{https://doi.org/10.1103/PhysRevB.97.064418}{Phys. Rev. B \textbf{97}, 064418 (2018)}.
\bibitem{Buess2004} M. Buess, R. H\"{o}llinger, T. Haug, K. Perzlmaier, U. Krey, D. Pescia, M. R. Scheinfein, D. Weiss, and C. H. Back, Fourier Transform Imaging of Spin Vortex Eigenmodes, \href{https://doi.org/10.1103/PhysRevLett.93.077207}{Phys. Rev. Lett. \textbf{93}, 077207 (2004)}.
\bibitem{Wang2017} Z. Wang, M. Li, and R. Wang, Resonance beyond frequency-matching: multidimensional resonance, \href{https://doi.org/10.1088/1367-2630/aa6275}{New J. Phys. \textbf{19}, 033012 (2017)}.
\bibitem{Kammerer2011} M. Kammerer, M. Weigand, M. Curcic, M. Noske, M. Sproll, A. Vansteenkiste, B. Van Waeyenberge, H. Stoll, G. Woltersdorf, C. H. Back, and G. Schuetz, Magnetic vortex core reversal by excitation of spin waves, \href{https://doi.org/10.1038/ncomms1277}{Nat. Commun. \textbf{2}, 279 (2011)}.
\bibitem{Wang2012} R. Wang and X. Dong, Sub-nanosecond switching of vortex cores using a resonant erpendicular magnetic field, \href{https://doi.org/10.1063/1.3687909}{Appl. Phys. Lett. \textbf{100}, 082402 (2012)}.
\bibitem{Guslienko2008} K. Y. Guslienko, A. N. Slavin, V. Tiberkevich, and S.-K. Kim, Dynamic Origin of Azimuthal Modes Splitting in Vortex-State Magnetic Dots, \href{https://doi.org/10.1103/PhysRevLett.101.247203}{Phys. Rev. Lett. \textbf{101}, 247203 (2008)}.
\bibitem{Schultheiss2019} K. Schultheiss, R. Verba, F. Wehrmann, K. Wagner, L. K\"{o}rber, T. Hula, T. Hache, A. K\'{a}kay, A. A. Awad, V. Tiberkevich, A. N. Slavin, J. Fassbender, and H. Schultheiss, Excitation of Whispering Gallery Magnons in a Magnetic Vortex, \href{https://doi.org/10.1103/PhysRevLett.122.097202}{Phys. Rev. Lett. \textbf{122}, 097202 (2019)}.
\bibitem{Wang2022} Z. Wang, H. Y. Yuan, Y. Cao, and P. Yan, Twisted Magnon Frequency Comb and Penrose Superradiance, \href{https://doi.org/10.1103/PhysRevLett.129.107203}{Phys. Rev. Lett. \textbf{129}, 107203 (2022)}.
\bibitem{Hoffmann2007} F. Hoffmann, G. Woltersdorf, K. Perzlmaier, A. N. Slavin, V. S. Tiberkevich, A. Bischof, D. Weiss, and C. H. Back, Mode degeneracy due to vortex core removal in magnetic disks, \href{https://doi.org/10.1103/PhysRevB.76.014416}{Phys. Rev. B \textbf{76}, 014416 (2007)}.
\bibitem{Salama2025} S. Salama, J.-V. Kim, A. Anane, and J.-P. Adam, Large frequency nonreciprocity of azimuthal spin-wave modes in submicron vortex state disks, \href{https://doi.org/10.1103/PhysRevB.111.134445}{Phys. Rev. B \textbf{111}, 134445 (2025)}.
\bibitem{Ivanov2002} B. A. Ivanov and G. M. Wysin, Magnon modes for a circular two-dimensional easy-plane ferromagnet in the cone state, \href{https://doi.org/10.1103/PhysRevB.65.134434}{Phys. Rev. B \textbf{65}, 134434 (2002)}.
\bibitem{Ivanov2005} B. A. Ivanov and C. E. Zaspel, High Frequency Modes in Vortex-State Nanomagnets, \href{https://doi.org/10.1103/PhysRevLett.94.027205}{Phys. Rev. Lett. \textbf{94}, 027205 (2005)}.
\bibitem{Uzunova2019} V. A. Uzunova and B. A. Ivanov, Magnon Modes for a Magnetic Disc in a Cone Vortex State, \href{https://doi.org/10.1063/1.5082327}{Low Temp. Phys. \textbf{45}, 92 (2019)}.
\bibitem{Uzunova2023} V. Uzunova, L. K\"{o}rber, A. Kavvadia, G. Quasebarth, H. Schultheiss, A. K\'{a}kay, and B. Ivanov, Nontrivial Aharonov-Bohm effect and alternating dispersion of magnons in cone-state ferromagnetic rings, \href{https://doi.org/10.1103/PhysRevB.108.174445}{Phys. Rev. B \textbf{108}, 174445 (2023)}.
\bibitem{Zakeri2010} Kh. Zakeri, Y. Zhang, J. Prokop, T.-H. Chuang, N. Sakr, W. X. Tang, and J. Kirschner, Asymmetric Spin-Wave Dispersion on Fe(110): Direct Evidence of the Dzyaloshinskii-Moriya Interaction, \href{https://doi.org/10.1103/PhysRevLett.104.137203}{Phys. Rev. Lett. \textbf{104}, 137203 (2010)}.
\bibitem{Di2015} K. Di, V. L. Zhang, H. S. Lim, S. C. Ng, M. H. Kuok, J. Yu, J. Yoon, X. Qiu, and H. Yang, Direct Observation of the Dzyaloshinskii-Moriya Interaction in a Pt/Co/Ni Film, \href{https://doi.org/10.1103/PhysRevLett.114.047201}{Phys. Rev. Lett. \textbf{114}, 047201 (2015)}.
\bibitem{Moon2013} J.-H. Moon, S.-M. Seo, K.-J. Lee, K.-W. Kim, J. Ryu, H.-W. Lee, R. D. McMichael, and M. D. Stiles, Spin-wave propagation in the presence of interfacial Dzyaloshinskii-Moriya interaction, \href{https://doi.org/10.1103/PhysRevB.88.184404}{Phys. Rev. B \textbf{88}, 184404 (2013)}.
\bibitem{Zingsem2019} B. W. Zingsem, M. Farle, R. L. Stamps, and R. E. Camley, Unusual nature of confined modes in a chiral system: Directional transport in standing waves, \href{https://doi.org/10.1103/PhysRevB.99.214429}{Phys. Rev. B \textbf{99}, 214429 (2019)}.
\bibitem{Cortes2013} D. Cort\'{e}s-Ortu\~{n}o and P. Landeros, Influence of the Dzyaloshinskii-Moriya interaction on the spin-wave spectra of thin films, \href{https://doi.org/10.1088/0953-8984/25/15/156001}{J. Phys.: Condens. Matter \textbf{25}, 156001 (2013)}.
\bibitem{Zhang2015} V. L. Zhang, K. Di, H. S. Lim, S. C. Ng, M. H. Kuok, J. Yu, J. Yoon, X. Qiu, and H. Yang, In-plane angular dependence of the spin-wave nonreciprocity of an ultrathin film with Dzyaloshinskii-Moriya interaction, \href{https://doi.org/10.1063/1.4926862}{Appl. Phys. Lett. \textbf{107}, 022402 (2015)}.
\bibitem{Wang2020} Z. Wang, Y. Cao, R. Wang, B. Liu, H. Meng, and P. Yan, Effect of Dzyaloshinskii-Moriya interaction on magnetic vortex switching driven by radial spin waves, \href{https://doi.org/10.1016/j.jmmm.2020.167014}{J. Magn. Magn. Mater. \textbf{512}, 167014 (2020)}.
\bibitem{Flores2020} C. Quispe Flores, C. Chalifour, J. Davidson, K. L. Livesey, and K. S. Buchanan, Semianalytical approach to calculating the dynamic modes of magnetic vortices with Dzyaloshinskii-Moriya interactions, \href{https://doi.org/10.1103/PhysRevB.102.024439}{Phys. Rev. B \textbf{102}, 024439 (2020)}.
\bibitem{Verba2020} R. V. Verba, D. Navas, S. A. Bunyaev, A. Hierro-Rodriguez, K. Y. Guslienko, B. A. Ivanov, and G. N. Kakazei, Helicity of magnetic vortices and skyrmions in soft ferromagnetic nanodots and films biased by stray radial fields, \href{https://doi.org/10.1103/PhysRevB.101.064429}{Phys. Rev. B \textbf{101}, 064429 (2020)}.
\bibitem{Vansteenkiste2014} A. Vansteenkiste, J. Leliaert, M. Dvornik, M. Helsen, F. Garcia-Sanchez, and B. Van Waeyenberge, The design and verification of MuMax3, \href{https://doi.org/10.1063/1.4899186}{AIP Adv. \textbf{4}, 107133 (2014)}.
\bibitem{Nembach2015} H. T. Nembach, J. M. Shaw, M. Weiler, E. Ju\'{e}, and T. J. Silva, Linear relation between Heisenberg exchange and interfacial Dzyaloshinskii-Moriya interaction in metal films, \href{https://doi.org/10.1038/nphys3418}{Nat. Phys. \textbf{11}, 825 (2015)}.
\bibitem{Kuepferling2023} M. Kuepferling, A. Casiraghi, G. Soares, G. Durin, F. Garcia-Sanchez, L. Chen, C. H. Back, C. H. Marrows, S. Tacchi, and G. Carlotti, Measuring interfacial Dzyaloshinskii-Moriya interaction in ultrathin magnetic films, \href{https://doi.org/10.1103/RevModPhys.95.015003}{Rev. Mod. Phys. \textbf{95}, 015003 (2023)}.
\bibitem{Chaurasiya2019} A. K. Chaurasiya, A. Kumar, R. Gupta, S. Chaudhary, P. K. Muduli, and A. Barman, Direct observation of unusual interfacial Dzyaloshinskii-Moriya interaction in graphene/NiFe/Ta heterostructures, \href{https://doi.org/10.1103/PhysRevB.99.035402}{Phys. Rev. B \textbf{99}, 035402 (2019)}.
\bibitem{Vas2024} J. V. Vas, R. Medwal, S. Manna, M. Mishra, A. Muller, J. R. Mohan, Y. Fukuma, M. Duchamp, and R. S. Rawat, Direct visualization of local magnetic domain dynamics in a 2D Van der Walls material/ferromagnet interface, \href{https://doi.org/10.1038/s42005-024-01861-w}{Commun. Phys. \textbf{7}, 407 (2024)}.
\bibitem{Cho2017} J. Cho, N.-H. Kim, S. K. Kang, H.-K. Hwang, J. Jung, H. J. M. Swagten, J.-S. Kim, and C.-Y. You, The sign of the interfacial Dzyaloshinskii-Moriya interaction in ultrathin amorphous and polycrystalline magnetic films, \href{https://doi.org/10.1088/1361-6463/aa89d4}{J. Phys. D: Appl. Phys. \textbf{50}, 425004 (2017)}.

\bibitem{Korber2020} L. K\"{o}rber, K. Schultheiss, T. Hula, R. Verba, J. Fassbender, A. K\'{a}kay, and H. Schultheiss, Nonlocal Stimulation of Three-Magnon Splitting in a Magnetic Vortex, \href{https://doi.org/10.1103/PhysRevLett.125.207203}{Phys. Rev. Lett. \textbf{125}, 207203 (2020)}.
\bibitem{Schryer1974} N. L. Schryer and L. R. Walker, The motion of $180^{\circ}$ domain walls in uniform dc magnetic fields, \href{https://doi.org/10.1063/1.1663252}{J. Appl. Phys. \textbf{45}, 5406 (1974)}.
\bibitem{Mulkers2017} J. Mulkers, B. Van Waeyenberge, and M. V. Milo\v{s}evi\'{c}, Effects of spatially engineered Dzyaloshinskii-Moriya interaction in ferromagnetic films, \href{https://doi.org/10.1103/PhysRevB.95.144401}{Phys. Rev. B \textbf{95}, 144401 (2017)}.
\bibitem{Wang2018} Z. Wang, B. Zhang, Y. Cao, and P. Yan, Probing the Dzyaloshinskii-Moriya Interaction via the Propagation of Spin Waves in Ferromagnetic Thin Films, \href{https://doi.org/10.1103/PhysRevApplied.10.054018}{Phys. Rev. Appl. \textbf{10}, 054018 (2018)}.
\bibitem{Guo2017} J. Guo, X. Zeng, and M. Yan, Spin-wave canting induced by the Dzyaloshinskii-Moriya interaction in ferromagnetic nanowires, \href{https://doi.org/10.1103/PhysRevB.96.014404}{Phys. Rev. B \textbf{96}, 014404 (2017)}.
\bibitem{Mimica2025} B. Mimica-Figari, P. Landeros, and R. A. Gallardo, Dzyaloshinskii-Moriya interaction and dipole-exchange curvature effects on the spin-wave spectra of magnetic nanotubes, \href{https://doi.org/10.1103/8b55-79md}{Phys. Rev. B \textbf{112}, 024421 (2025)}.
\bibitem{Yu2016} W. Yu, J. Lan, R. Wu, and J. Xiao, Magnetic Snell's law and spin-wave fiber with Dzyaloshinskii-Moriya interaction, \href{https://doi.org/10.1103/PhysRevB.94.140410}{Phys. Rev. B \textbf{94}, 140410 (2016)}.
\end{thebibliography}
\end{document}